\documentclass[aps,nofootinbib,preprintnumbers,superscriptaddress]{revtex4}

\usepackage{textcomp}
\usepackage{gensymb}

\usepackage{mathtools}
\usepackage{amsfonts}
\usepackage{amsmath}
\usepackage{multirow}

\usepackage{subcaption}

\usepackage[paperwidth=210mm,paperheight=297mm,centering,hmargin=2cm,vmargin=2.5cm]{geometry}

\usepackage{hyperref}

\allowdisplaybreaks

\def\bfs{\mathbf{1}}
\def\bfsp{\mathbf{1^\prime}}
\def\bfspp{\mathbf{1^{\prime\prime}}}
\def\bft{\mathbf{3}}

\pdfoutput=1
\begin{document}

\title{Two A4 modular symmetries for Tri-Maximal 2 mixing}

\author{Ivo de Medeiros Varzielas}
\email{ivo.de@udo.edu}
\affiliation{CFTP, Departamento de F\'{\i}sica, Instituto Superior T\'{e}cnico,\\
Universidade de Lisboa, Avenida Rovisco Pais 1, 1049 Lisboa, Portugal\\
}
\author{Jo\~ao Louren\c{c}o}
\email{joao.freitas.lourenco@tecnico.ulisboa.pt}
\affiliation{CFTP, Departamento de F\'{\i}sica, Instituto Superior T\'{e}cnico,\\
Universidade de Lisboa, Avenida Rovisco Pais 1, 1049 Lisboa, Portugal\\
}
\begin{abstract}
We construct a lepton flavour model based on two $A_4$ modular symmetries. The two $A_4$ are broken by a bi-triplet field to the diagonal $A_4$ subgroup, resulting in an effective modular $A_4$ flavour symmetry with two moduli. We employ these moduli as stabilisers, that preserve distinct residual symmetries, enabling us to obtain Tri-Maximal 2 (TM2) mixing with a minimal field content,
flavonless at the effective scale, below the breaking to the single $A_4$.
\end{abstract}

\maketitle
\section{Introduction \label{intro}}

Flavour symmetries, both discrete and continuous, have been extensively employed in the literature as a way to solve the puzzling questions associated with flavour. Examples of well-known discrete symmetries applied to flavour are $S_3$, $A_4$, $S_4$ and $A_5$.
Recently, these same symmetries are used in flavour models as modular symmetries \cite{deAdelhartToorop:2011re, Feruglio:2017spp}: $\Gamma_2\simeq S_3$~\cite{Kobayashi:2018vbk, Kobayashi:2018wkl, Okada:2019xqk, Mishra:2020gxg, Du:2020ylx}, 
$\Gamma_3\simeq A_4$~
\cite{Feruglio:2017spp, Criado:2018thu, Kobayashi:2018scp, Okada:2018yrn, Novichkov:2018yse, Nomura:2019jxj, Nomura:2019yft, Ding:2019zxk, Zhang:2019ngf, Okada:2019mjf, Nomura:2019lnr, Asaka:2019vev, Nomura:2019xsb, Kobayashi:2019gtp, Wang:2019xbo, Okada:2020dmb, Ding:2020yen, Behera:2020sfe, Nomura:2020opk, Nomura:2020cog, Behera:2020lpd, Asaka:2020tmo, Nagao:2020snm, Hutauruk:2020xtk},
$\Gamma_4\simeq S_4$~\cite{Penedo:2018nmg, Novichkov:2018ovf, deMedeirosVarzielas:2019cyj, King:2019vhv, Kobayashi:2019mna, Okada:2019lzv, Kobayashi:2019xvz, Wang:2019ovr, Wang:2020dbp}, 
and
$\Gamma_5\simeq A_5$~\cite{Novichkov:2018nkm, Ding:2019xna}.
Recently, $\Gamma_7 \simeq PSL(2,\mathbb{Z}_7)$ was studied  \cite{Ding:2020msi},
and
\cite{Gehrlein:2020jnr} studied the mass sum rules arising in these models.

Models using multiple $S_4$ modular symmetries can be found at \cite{deMedeirosVarzielas:2019cyj, King:2019vhv}, where a general mechanism of employing multiple modular symmetries is presented. In this framework of multiple modular symmetries, each one with its own modulus field (one for the charged leptons, one or two for the neutrinos), a high energy theory with multiple symmetry groups is constructed, which is then broken to a low energy model with a single modular symmetry, thus effectively explaining the use of multiple moduli together with a single modular symmetry group \footnote{An alternative approach to incorporate several moduli was proposed in \cite{Ding:2020zxw}.}. This low energy model with a single modular symmetry is also broken when their modulus fields gain different vacuum expectation values (VEV's) at fixed points of the modular symmetry (stabilisers) \footnote{We note that addressing the mechanism for giving the modulus fields their VEV's, known as the moduli stabilization problem, is beyond the scope of the present work, see \cite{Novichkov:2022wvg}.}. The preserved residual symmetries then lead to the realization of different mass textures in the charged lepton and neutrino sectors in modular flavour models without flavons.
These modular symmetries are thus able to generate all masses and mixing parameters for the leptons, using a much smaller set of free parameters.
As an example, a $S_4$ flavour model featuring TM1 mixing \cite{Albright:2008rp, Varzielas:2012pa} is constructed in an elegant manner from multiple $S_4$ modular symmetries \cite{deMedeirosVarzielas:2019cyj}.

Before the mixing angles were observed experimentally with more precision, a commonly used mixing texture for the PMNS matrix was the tri-bimaximal mixing (TBM):
\begin{equation}
U_{TBM}=\begin{pmatrix}
\sqrt{\frac{2}{3}} & \sqrt{\frac{1}{3}} & 0\\
-\sqrt{\frac{1}{6}} & \sqrt{\frac{1}{3}} & -\sqrt{\frac{1}{2}}\\
-\sqrt{\frac{1}{6}} & \sqrt{\frac{1}{3}} & \sqrt{\frac{1}{2}}
\end{pmatrix}.
\label{TBM matrix}
\end{equation}
This ansatz, ruled out since the measurement of non-zero $\theta_{13}$ mixing angle, remains an appealing leading order solution with no free parameters. Mixing schemes such as Tri-Maximal 1 (TM1) \cite{Albright:2008rp} and Tri-Maximal 2 (TM2) \cite{Grimus:2008tt} preserve respectively the 1st and 2nd columns of tri-bimaximal mixing \cite{Albright:2010ap}, and remain viable.

In this paper, we will use two $A_4$ modular symmetries to build a model that leads to TM2 mixing due to two independent modulus stabilisers, similarly to the use of multiple $S_4$ modular symmetries in \cite{King:2019vhv,deMedeirosVarzielas:2019cyj}. We note that \cite{Novichkov:2018yse} employs a single $A_4$ modular symmetry and two moduli in a model leading to TM2 mixing, where neutrino masses arise through the effective Weinberg operator. 
In the model constructed here, we use the type I seesaw mechanism and justify the presence of two distinct moduli by starting with two $A_4$ symmetries which are subsequently broken to the diagonal subgroup.

In Section \ref{mms} we briefly review the framework of multiple modular symmetries. In Section \ref{sec:A4res} we describe the $A_4$ modular symmetry and respective stabilisers associated to residual symmetries. In Section \ref{sec:models A4} we present one possible model for TM2 mixing. We conclude in Section \ref{sec:conclusion}.

\section{Flavour modular symmetries \label{mms}}

This section provides the definition of modular group and modular forms, and some fundamental aspects of constructing a realistic model with single and multiple modular symmetries, as in \cite{deMedeirosVarzielas:2019cyj}.

\subsection{Modular group and modular forms}

The modular group $\overline{\Gamma}$ is the group of linear fractional transformations $\gamma$ that act on the complex modulus $\tau$, for $\tau$ in the upper-half complex plane, i.e. $Im(\tau)>0$:
\begin{equation}
\gamma:\tau\rightarrow\gamma\tau=\frac{a\tau+b}{c\tau+d},
\end{equation}
where $a,b,c,d$ are integers and satisfy $ad-bc=1$. 

It is convenient to use $2 \times 2$ matrices to represent the elements of $\overline{\Gamma}$ as
\begin{equation}
\overline{\Gamma}=\left\{ \begin{pmatrix}
a&b\\
c&d
\end{pmatrix}/\{\pm 1\},~a,b,c,d\in \mathbb{Z}, ~ ad-bc=1 \right\}.
\label{modular group matrices}
\end{equation}

Note that, since $\gamma$ and $-\gamma$ are the same modular transformation, the  group $\overline{\Gamma}$ is isomorphic to $PSL(2,\mathbb{Z}) = SL(2,\mathbb{Z})/\mathbb{Z}_2$, where $SL(2,\mathbb{Z})$ is the group of $2 \times 2$ matrices with integer entries and determinant one.

The modular group has two generators, $S_{\tau}$ and $T_{\tau}$, which satisfy $S_{\tau}^2=(S_{\tau}T_{\tau})^3=1$. One possible choice for these generators is the following:
\begin{equation}
S_{\tau}:\tau\rightarrow-\frac{1}{\tau},~T_{\tau}:\tau\rightarrow\tau+1
\end{equation}
and their corresponding representations are
\begin{equation}
S_{\tau}=\begin{pmatrix}
0&1\\
-1&0
\end{pmatrix},~
T_{\tau}=\begin{pmatrix}
1&1\\
0&1
\end{pmatrix}.
\label{generators representation 2times2}
\end{equation}

It is possible to define subgroups $\overline{\Gamma}(N)$ of $\overline{\Gamma}$ modding out the entries of the representation matrices:
\begin{equation}
\overline{\Gamma}(N)=\left\{ \begin{pmatrix}
a&b\\
c&d
\end{pmatrix}\in PSL(2,\mathbb{Z}),~
\begin{pmatrix}
a&b\\
c&d
\end{pmatrix}=
\begin{pmatrix}
1&0\\
0&1
\end{pmatrix} ~(mod~\text{N})\right\}.
\end{equation}

Although the groups $\overline{\Gamma}(N)$ are discrete but infinite, the quotient groups $\Gamma_{N}=\overline{\Gamma}/\overline{\Gamma}(N)$ are finite, thus being called finite modular groups. For $N \leq 5$, these groups are isomorphic to well-known groups: 
$\Gamma_2 \simeq S_3$,
$\Gamma_3 \simeq A_4$,
$\Gamma_4 \simeq S_4$,
$\Gamma_5 \simeq A_5$.
These finite modular groups can be obtained by imposing an additional condition, $T_{\tau}^N=1$, which implies that $\tau=\tau+N$.

Modular forms of weight $2k$ and level $N$ are holomorphic functions of $\tau$ that transform
under $\overline{\Gamma}(N)$ in the following way:
\begin{equation}
f(\gamma\tau)=(c\tau+d)^{2k}f(\tau),~\gamma=
\begin{pmatrix}
a&b\\c&d
\end{pmatrix}
\in \overline{\Gamma}(N),
\end{equation}
where $k$ is a non-negative integer and $N$ is natural (we are only interested in even weights).
These modular forms are invariant under $\overline{\Gamma}(N)$,
up to the factor $(c\tau + d)^{2k}$, but they transform under the quotient group
$\Gamma_N$. 

Modular forms of weight $2k$ and level $N$ span a linear space of finite dimension $\mathcal{M}_{2k}(\overline{\Gamma}(N))$. Here we are interested in the $A_4$ modular group, so it is useful to know that the dimension of the linear space of modular forms of weight $2k$ and level $3$ is $2k+1$.
It is possible to choose a basis in $\mathcal{M}_{2k}(\overline{\Gamma}(N))$ such
that the transformation of the modular forms under $\Gamma_N$ is described by a unitary representation $\rho$ of $\Gamma_N$:
\begin{equation}
f_i(\gamma\tau)=(c\tau+d)^{2k}\rho(\tilde{\gamma})_{ij}f_j(\tau),~
\tilde{\gamma}
\in\text{equivalence class of }\gamma\in\bar{\Gamma}\text{ in }\Gamma_N.
\end{equation}

\subsection{Models with a single modular symmetry}

Considering an N = 1 supersymmetric model invariant under one full modular group, the action in general takes the form
\begin{equation}
S=\int 
d^4x d^2\theta d^2\overline{\theta}~
K(\phi_i,\overline{\phi}_i;\tau,\overline{\tau})+\left(
\int d^4xd^2\theta~W(\phi_i;\tau)+h.c.\right).
\end{equation}
Under $\bar{\Gamma}$ the K\"ahler potential $K$ transforms at most by a K\"ahler transformation and the superpotential $W$ stays invariant:
\begin{align}
K(\phi_i,\overline{\phi}_i;\tau,\overline{\tau}) & \rightarrow 
K(\phi_i,\overline{\phi}_i;\tau,\overline{\tau})+f(\phi_i;\tau)+\overline{f}(\overline{\phi}_i; \overline{\tau})\\
W(\phi_i;\tau) & \rightarrow W(\phi_i;\tau).
\end{align}
We assume here the minimal form of the K\"ahler potential. The effects of considering non-minimal forms of the K\"ahler potential may be relevant and are discussed in \cite{Chen:2019ewa, Feruglio:2021dte}.

The superpotential is in general a function of the modulus $\tau$ and superfields $\phi_i$ and can be expanded as:
\begin{equation}
W(\phi_i;\tau)=\sum_n \sum_{\{i_1,...,i_n\}} \sum_{I_Y} ~
(Y_{I_Y}\phi_{i_1}...\phi_{i_n})_{\mathbf{1}}.
\end{equation}

We want the superpotential to be invariant under $\bar{\Gamma}$. This is possible if we assume the couplings $Y_{I_Y}$ to be multiplet modular forms, and the superfields $\phi_i$ to transform as
\begin{align}
\phi_i(\tau) & \rightarrow \phi_i(\gamma\tau)=(c\tau+d)^{-2k_i}\rho_{I_i}(\gamma)\phi_i(\tau)\\
Y_{I_Y}(\tau) & \rightarrow 
Y_{I_Y}(\gamma\tau)=(c\tau+d)^{2k_Y}\rho_{I_Y}(\gamma)Y_{I_Y}(\tau),
\end{align}
where $-2k_i$ is the modular weight of $\phi_i$, $I_i$ labels the irrep of $\Gamma_N$ under which $\phi_i$ transforms, $2k_Y$ is the modular weight of $Y_{I_Y}$, $I_Y$ labels the irrep of $\Gamma_N$ under which $Y_{I_Y}$ transforms and $\rho_{I_i}(\gamma)$ and $\rho_{I_Y}(\gamma)$ are the unitary representation matrices of $\gamma\in\Gamma_N$.
For the superpotential to be invariant as wanted, the sum of the weights needs to equal zero, i.e. $k_Y=k_{i_1}+\ldots+k_{i_n}$, and the multiplication of the representations $I_Y \times I_{i_1} \times\ldots\times I_{i_n}$ has to contain an invariant singlet.

\subsection{Models with multiple modular symmetries}

It is also possible to consider a theory that has multiple modular symmetries, based on a series of M modular groups 
$\overline{\Gamma}^1$, 
$\overline{\Gamma}^2$, $\ldots$, $\overline{\Gamma}^M$, where the modulus field for each symmetry $\overline{\Gamma}^J$, $J=1,\ldots,M$, is denoted as $\tau_J$. The associated modular transformations take the form:
\begin{equation}
\gamma_J:\tau_J\rightarrow\gamma_J\tau_J=
\frac{a_J\tau_J+b_J}{c_J\tau_J+d_J}.
\end{equation}

A series of finite modular groups $\Gamma_{N_J}^J$ for $J = 1,\ldots,M$ can be obtained by modding out
an integer $N_J$ as done for only one modular group in the previous subsection and taking the quotient finite groups. Take into account that $N_J$ does not need to be identical to $N_{J'}$ for $J \neq J'$.

Consider again an N = 1 supersymmetric model invariant under multiple full modular groups; the action is
extended to the form
\begin{equation}
S=\int 
d^4x d^2\theta d^2\overline{\theta}~
K(\phi_i,\overline{\phi}_i;\tau_1,\ldots,\tau_M,\overline{\tau}_1,\ldots,\overline{\tau}_M)+\left(
\int d^4xd^2\theta~W(\phi_i;\tau_1,\ldots,\tau_M)+h.c.\right).
\end{equation}

In this case, the superpotential expansion in powers of the superfields takes the form
\begin{equation}
W(\phi_i;\tau_1,\ldots,\tau_M)=\sum_n \sum_{\{i_1,\ldots,i_n\}} \sum_{(I_{Y,1},\ldots,I_{Y,M})}~
(Y_{(I_{Y,1},\ldots,I_{Y,M})}\phi_{i_1}\ldots\phi_{i_n})_{\mathbf{1}}
\end{equation}
and under any finite modular transformation $\gamma_1,\ldots,\gamma_M$ in $\Gamma^1_{N_1}\times \Gamma^2_{N_2} \times \ldots \times \Gamma^M_{N_M}$ the superfields and multiplet couplings transform as
\begin{align}
\phi_i(\tau_1,\ldots,\tau_M) & \rightarrow \phi_i(\gamma_1\tau_1,\ldots,\gamma_M\tau_M)\nonumber\\
=&\prod_{J=1,\ldots,M} (c_J\tau_J+d_J)^{-2k_{i,J}}
\bigotimes_{J=1,\ldots,M}\rho_{I_{i,J}}(\gamma_J)~\phi_i(\tau_1,\ldots,\tau_M)\\
Y_{(I_{Y,1},\ldots,I_{Y,M})}(\tau_1,\ldots,\tau_M) & \rightarrow
Y_{(I_{Y,1},\ldots,I_{Y,M})}(\gamma_1\tau_1,\ldots,\gamma_M\tau_M)\nonumber\\
=&\prod_{J=1,\ldots,M}
(c_J\tau_J+d_J)^
{2k_{Y,J}}
\bigotimes_{J=1,\ldots,M}\rho_{I_{Y,J}}(\gamma_J)~Y_{(I_{Y,1},\ldots,I_{Y,M})}(\tau_1,\ldots,\tau_M).
\label{transform superfields and couplings for multi fields}
\end{align}
As discussed previously, for the superpotential to be invariant, $k_{Y,J}=k_{i_1,J}+\ldots+k_{i_n,J}$, and $I_{Y,J} \times I_{i_1,J} \times\ldots\times I_{i_n,J}$ must contain an invariant singlet, for $J=1,\ldots,M$.

\section{Modular \texorpdfstring{$A_4$}{A4} symmetry and residual symmetries \label{sec:A4res}}

The model present in this paper uses two $A_4$ modular symmetries, so the following subsection will deal with the $A_4$ symmetry group in particular. The stabilisers which are included here apply for the specific case of $A_4$ modular symmetries and, as well as the stabilisers for the modular groups from $N=2$ to $5$, can be found in \cite{deMedeirosVarzielas:2020kji} (we note also the stabilizers or fixed points for $N=3,4$ were presented in \cite{Ding:2019gof}). The directions at the stabilisers can also be found in \cite{Novichkov:2018yse}, although the factors for the modular forms were corrected here.

\subsection{Modular \texorpdfstring{$A_4$}{A4} symmetry and modular forms of level 3}

The group $A_4$ is the group of even permutations of 4 objects and has 12 elements. It is generated by two operators $S_{\tau}$ and $T_{\tau}$ obeying
\begin{equation}
S_{\tau}^2=(S_{\tau}T_{\tau})^3=T_{\tau}^3=1.
\end{equation}

This group has three singlets and one triplet as its irreducible representations and the multiplication rules and other properties can be found in Appendix \ref{app:A4}.
In the so-called complex basis (basis where $T_{\tau}$ is diagonal), the $A_4$ generators in the triplet representation are
\begin{equation}
\rho_{\bft}(S_{\tau})=\frac{1}{3}
\begin{pmatrix}
-1&2&2\\
2&-1&2\\
2&2&-1
\end{pmatrix}~~\text{and}~~
\rho_{\bft}(T_{\tau})=
\begin{pmatrix}
1&0&0\\
0&\omega&0\\
0&0&\omega^2
\end{pmatrix}.~~\omega=e^{i2\pi/3}.
\label{triplet representations A4}
\end{equation}

The flavour model we build here employ $A_4$ as a modular symmetry group and the Yukawa couplings are going to be modular forms.
The three linearly independent weight 2 modular forms of level 3, $Y_\bft^{(2)}=(Y_1,Y_2,Y_3)$, form a triplet of $A_4$ and can be expressed in terms of the Dedekind eta functions (see Appendix \ref{modular forms A4}).
The modular forms of higher weight can be generated starting from these modular forms of weight 2.
For example, the five linearly independent weight 4 modular forms decompose into a triplet $\bft$ and two singlets $\bfs$ and $\bfsp$. 
Using the weight 2 modular forms, one obtains the weight 4 modular forms:
\begin{equation}
Y_{\bft}^{(4)}
=\frac{2}{3}
\begin{pmatrix}
Y_1^2-Y_2Y_3\\
Y_3^2-Y_1Y_2\\
Y_2^2-Y_1Y_3
\end{pmatrix}
\end{equation}
and 
\begin{equation}
Y_{\bfs}^{(4)}=
Y_1^2+2Y_2Y_3,~~
Y_{\bfs'}^{(4)}=
Y_3^2+2Y_1Y_2.
\end{equation}
The singlet $\bfspp$ vanishes because $Y_\bft^{(2)}(\tau)$ satisfy the constraint
\begin{equation}
Y_{\bfs''}^{(4)}=Y_2^2+2Y_1Y_3=0.
\end{equation}
Note that here a factor $2/3$ was included in the definition for $Y_{\bft}^{(4)}$ in accordance with \cite{Novichkov:2018yse} although no such factor is present in \cite{Feruglio:2017spp}.

Furthermore, the modular forms of weight 6, whose linear space has dimension 7 and decomposes into 2 triplets and 1 singlet, are \cite{Feruglio:2017spp}:
\begin{align}
Y_{{\bft_1}}^{(6)}&=
\begin{pmatrix}
Y_1^3+2Y_1Y_2Y_3\\
Y_1^2Y_2+2Y_2^2Y_3\\
Y_1^2Y_3+2Y_3^2Y_2
\end{pmatrix}\\
Y_{{\bft_2}}^{(6)}&=
\begin{pmatrix}
Y_3^3+2Y_1Y_2Y_3\\
Y_3^2Y_1+2Y_1^2Y_2\\
Y_3^2Y_2+2Y_2^2Y_1
\end{pmatrix}
\end{align}
and 
\begin{equation}
Y_{\bfs}^{(6)}=
Y_1^3+Y_2^3+Y_3^3-3Y_1Y_2Y_3,
\end{equation}
and again the other triplet that we are able to construct vanishes:
\begin{equation}
Y_{{\bft_3}}^{(6)}=
\begin{pmatrix}
Y_2^3+2Y_1Y_2Y_3\\
Y_2^2Y_3+2Y_3^2Y_1\\
Y_2^2Y_1+2Y_1^2Y_3
\end{pmatrix}=0.
\end{equation}
For completeness, the linear space of the modular forms with weight 8 has dimension 9 and decomposes into three singlets, the first invariant:
\begin{align}
\label{eq:mf singlet 8} Y_{\bfs}^{(8)}&=Y_1^4+4Y_1^2Y_2Y_3 + 4Y_2^2Y_3^2\\
\label{eq:mf singletp 8} Y_{\bfsp}^{(8)}&=2Y_1^3Y_2 + 4Y_1Y_2^2Y_3 + Y_1^2Y_3^2 + 2Y_2Y_3^3\\
\label{eq:mf singletpp 8} Y_{\bfspp}^{(8)}&=Y_3^4+4Y_1Y_2Y_3^2 + 4Y_1^2Y_2^2
\end{align}
and two triplets:
\begin{align}
\label{eq:mf triplet1 8} Y_{{\bft_1}}^{(8)} & =
\begin{pmatrix}
Y_1^4+Y_1^2Y_2Y_3 - 2Y_2^2Y_3^2\\
Y_1^2Y_3^2 - Y_1^3Y_2 - 2 Y_1Y_2^2Y_3 + 2 Y_2Y_3^3\\
Y_1^2Y_2^2 - Y_1^3Y_3 - 2 Y_1Y_2Y_3^2 + 2 Y_2^3Y_3
\end{pmatrix}\\
\label{eq:mf triplet2 8} Y_{{\bft_2}}^{(8)} & =
\begin{pmatrix}
Y_1^4 + Y_1Y_2^3 - 3Y_1^2Y_2Y_3 + Y_1Y_3^3 \\
Y_2^4 + Y_2Y_3^3 - 3Y_1Y_2^2Y_3 + Y_2Y_1^3 \\
Y_3^4 + Y_3Y_2^3 - 3Y_1Y_2Y_3^2 + Y_3Y_1^3
\end{pmatrix}.
\end{align}

\subsection{Stabilisers and residual symmetries of modular \texorpdfstring{$A_4$}{A4}}

The stabilisers of the symmetry play a crucial role in preserving residual symmetries. Given an element $\gamma$ in the modular group $A_4$, a stabiliser $\tau_\gamma$ of $\gamma$ corresponds to a fixed point in the upper half complex plane that transforms as $\gamma\tau_{\gamma}=\tau_{\gamma}$. Once the modular field acquires a VEV at this special point, 
$\langle\tau\rangle=\tau_{\gamma}$, the modular symmetry is broken but an Abelian residual modular symmetry generated by $\gamma$ is preserved. Obviously, acting $\gamma$ on the modular form at its stabiliser leaves the modular form invariant:
\begin{equation}
\gamma:Y_I(\tau_{\gamma})\rightarrow Y_I(\gamma\tau_{\gamma})=
Y_I(\tau_{\gamma}),
\end{equation}
which implies that
\begin{equation}
\rho_I(\gamma)Y_I(\tau_{\gamma})=
(c\tau_{\gamma}+d)^{-2k}Y_I(\tau_{\gamma}).
\end{equation}

This means that, at the stabiliser, the modular form is an eigenvector of the representation matrix $\rho_\bft(\gamma)$ for the given stabiliser that corresponds to the eigenvalue $(c\tau_{\gamma}+d)^{-2k}$, and thus the directions of the modular forms at the stabilisers can be easily determined.
Furthermore, since the representation matrix is unitary, $|c\tau_{\gamma}+d|=1$. 

The stabilisers for the $A_4$ modular group are shown in TABLE \ref{table:stabilisers A4} \cite{deMedeirosVarzielas:2020kji}. 
\begin{table}[ht]
\centering
\begin{tabular}{c|c} 
\toprule
$\gamma$ & $\tau_{\gamma}$\\ [0.5ex] 
\colrule
$T_{\tau}, T_{\tau}^2$ & $i\infty,~ \frac{3}{2}+\frac{i}{2\sqrt{3}}$ \\
$S_{\tau}T_{\tau}, T_{\tau}^2S_{\tau}$ & $1,~ -\frac{1}{2}+\frac{i\sqrt{3}}{2}$ \\
$T_{\tau}S_{\tau}T_{\tau}, S_{\tau}T_{\tau}S_{\tau}$ & $0,~ \frac{3}{2}+\frac{i\sqrt{3}}{2}$ \\ 
$T_{\tau}S_{\tau}, S_{\tau}T_{\tau}^2$ & $-1,~ \frac{1}{2}+\frac{i\sqrt{3}}{2}$ \\
$T_{\tau}^2S_{\tau}T_{\tau}$ & $-1+i,~ \frac{1}{2}+\frac{i}{2}$ \\
$S_{\tau}$ & $i,~ \frac{3}{2}+\frac{i}{2}$ \\  
$T_{\tau}S_{\tau}T_{\tau}^2$ & $1+i,~ -\frac{1}{2}+\frac{i}{2}$ \\ [1ex] 
\botrule
\end{tabular}
\caption{Stabilisers for the $A_4$ elements \cite{deMedeirosVarzielas:2020kji}.}
\label{table:stabilisers A4}
\end{table}

For the transformations $S_{\tau}$, $T_{\tau}$, $S_{\tau}T_{\tau}$ and $T_{\tau}S_{\tau}$, the coefficients $(c\tau_{\gamma}+d)^{-2k}$ are
\begin{equation}
(c \tau_{\gamma} + d)^{-2k} = 
\left\{
\begin{array}{ll}
      (-1)^{k} & \gamma = S_{\tau}, \tau_{S_{\tau 1}} = i \text{ or } \tau_{S_{\tau 2}} = \frac{3}{2} + \frac{i}{2} \\
      1 & \gamma = T_{\tau}, \tau_{T_{\tau 1}} = i\infty \\
      \omega^{2k} & \gamma = T_{\tau}, \tau_{T_{\tau 2}} = \frac{3}{2} + \frac{i}{2\sqrt{3}} \\
      \omega^{2k} & \gamma = S_{\tau}T_{\tau}, \tau_{S_{\tau}T_{\tau}} = \omega \\
      \omega^{2k} & \gamma = T_{\tau}S_{\tau}, \tau_{T_{\tau}S_{\tau}} = -\omega^2 \\
\end{array}
\right. .
\label{stabilisers eigenvalues A4}
\end{equation}

The directions of the modular forms of weight $2k=2$, $4$, $6$ and $8$ for the stabilisers of these four elements are shown in TABLE \ref{table:directions A4}. Additionally, we include the factors for each modular form (\cite{Novichkov:2018yse} presents the directions for the modular forms of weight 2 and 4, although the factors have been corrected here for the weight 4 triplets). These factors are written in function of $Y$, which is defined in general as the first component $Y_1$ of $Y_\bft^{(2)}$, except for $\tau_{T_{\tau 2}}=\frac{3}{2}+\frac{i}{2\sqrt{3}}$, when we define it as the third component $Y_3$ of that triplet since the first component happens to vanish. For $Y$, the definitions for the weight 2 modular forms present in Appendix \ref{modular forms A4} were used. The values the modular form singlets of weight 4, 6 and 8 take at the stabilisers are also included in TABLE \ref{table:singlets A4}.

The other two stabilisers for $S_{\tau}T_{\tau}$ and $T_{\tau}S_{\tau}$ were not considered since the modular forms approach infinity for these two values of the modulus field. 
Notice that the two stabilisers of $S_\tau$ and $T_\tau$ stabilise these modular transformations but for different, although equivalent, representations in terms of $2\times2$ matrices, and thus for different $c$ and $d$. This explains the different eigenvalues obtained for the two stabilisers of $T_\tau$. However, for $S_\tau$ the eigenvalues happen to be the same for both stabilisers. Nevertheless, the directions for these two stabilisers of $S_\tau$ are indeed different. In this case the difference comes from the existence of two eigenvectors for the same eigenvalue, eigenvectors that are introduced in the example that follows.

\begin{table}[t!]
\resizebox{.98\hsize}{!}{
\centering
{\renewcommand{\arraystretch}{1.2}
\begin{tabular}{c|c|c|c|c|c|c}
\toprule
\multirow{2}{*}{$\tau_\gamma$} & weight 2 & weight 4 & \multicolumn{2}{c|}{weight 6} & \multicolumn{2}{c}{weight 8}\\
\cline{2-7} 
& $\bft$ & $\bft$ & $\bft_1$ & $\bft_2$ & $\bft_1$ & $\bft_2$ \\
\colrule
\multicolumn{1}{l|}{$\tau_{S_{\tau}T_{\tau}}=-\frac{1}{2}+\frac{i\sqrt{3}}{2}$} & $Y\begin{pmatrix}1\\\omega\\-\frac{1}{2}\omega^2\end{pmatrix}$ & $Y^2\begin{pmatrix}1\\-\frac{1}{2}\omega\\\omega^2\end{pmatrix}$ & $0$ & $-\frac{9}{8} Y^3 \begin{pmatrix}1\\-2\omega\\-2\omega^2\end{pmatrix}$ & $0$ & $\frac{27}{8} Y^4 \begin{pmatrix}1\\ \omega \\ -\frac{1}{2} \omega^2 \end{pmatrix}$\\ 
\multicolumn{1}{l|}{$\tau_{T_{\tau}S_{\tau}}=\frac{1}{2}+\frac{i\sqrt{3}}{2}$} & $Y\begin{pmatrix}1\\\omega^2\\-\frac{1}{2}\omega\end{pmatrix}$ & $Y^2\begin{pmatrix}1\\-\frac{1}{2}\omega^2\\\omega\end{pmatrix}$ & $0$ & $-\frac{9}{8} Y^3 \begin{pmatrix}1\\-2\omega^2\\-2\omega\end{pmatrix}$ & $0$ & $\frac{27}{8} Y^4 \begin{pmatrix}1\\ \omega^2 \\ -\frac{1}{2} \omega \end{pmatrix}$ \\
\multicolumn{1}{l|}{$\tau_{S_{\tau 1}}=i$} & $Y\begin{pmatrix}1\\1-\sqrt{3}\\-2+\sqrt{3}\end{pmatrix}$ & $(4-2\sqrt{3})Y^2\begin{pmatrix}1\\1\\1\end{pmatrix}$ & $(6\sqrt{3}-9)Y^3\begin{pmatrix}1\\1-\sqrt{3}\\-2+\sqrt{3}\end{pmatrix}$ & $(21\sqrt{3}-36)Y^3\begin{pmatrix}1\\-2-\sqrt{3}\\1+\sqrt{3}\end{pmatrix}$ & $(63\sqrt{3}-108)Y^4\begin{pmatrix}1\\1\\1\end{pmatrix}$ & $0$ \\
\multicolumn{1}{l|}{$\tau_{S_{\tau 2}}=\frac{3}{2}+\frac{i}{2}$} & $Y\begin{pmatrix}1\\1+\sqrt{3}\\-2-\sqrt{3}\end{pmatrix}$ & $(4+2\sqrt{3})Y^2\begin{pmatrix}1\\1\\1\end{pmatrix}$ & $-(6\sqrt{3}+9)Y^3\begin{pmatrix}1\\1+\sqrt{3}\\-2-\sqrt{3}\end{pmatrix}$ & $-(21\sqrt{3}+36)Y^3\begin{pmatrix}1\\-2+\sqrt{3}\\1-\sqrt{3}\end{pmatrix}$ & $-(63\sqrt{3}+108)Y^4\begin{pmatrix}1\\1\\1\end{pmatrix}$ & $0$ \\
\multicolumn{1}{l|}{$\tau_{T_{\tau 1}}=i\infty$} & $Y\begin{pmatrix}1\\0\\0\end{pmatrix}$ & $\frac{2}{3} Y^2\begin{pmatrix}1\\0\\0\end{pmatrix}$ & $Y^3\begin{pmatrix}1\\0\\0\end{pmatrix}$ & $0$ & $Y^4\begin{pmatrix}1\\0\\0\end{pmatrix}$ & $Y^4\begin{pmatrix}1\\0\\0\end{pmatrix}$ \\
\multicolumn{1}{l|}{$\tau_{T_{\tau 2}}=\frac{3}{2}+\frac{i}{2\sqrt{3}}$} & $Y\begin{pmatrix}0\\0\\1\end{pmatrix}$ & $\frac{2}{3} Y^2\begin{pmatrix}0\\1\\0\end{pmatrix}$ & $0$ & $Y^3\begin{pmatrix}1\\0\\0\end{pmatrix}$ & $0$ & $Y^4\begin{pmatrix}0\\0\\1\end{pmatrix}$ \\
\botrule
\end{tabular}}}
\caption{Directions for the modular forms of weight 2, 4, 6 and 8 of level 3 for four $A_4$ elements (Y in TABLE \ref{table:singlets A4}).}
\label{table:directions A4}
\end{table}

\begin{table}[t!]
\resizebox{.98\hsize}{!}{
\centering
{\renewcommand{\arraystretch}{1.2}
\begin{tabular}{c|c|c|c|c|c|c|c}
\toprule 
\multirow{2}{*}{$\tau_\gamma$} & \multicolumn{2}{c|}{weight 4} & weight 6 & \multicolumn{3}{c|}{weight 8} & \multirow{2}{*}{$Y$}\\
\cline{2-7} 
& $\bfs$ & $\bfsp$ & $\bfs$ & $\bfs$ & $\bfsp$ & $\bfspp$ & \\
\colrule
\multicolumn{1}{l|}{$\tau_{S_{\tau}T_{\tau}}=-\frac{1}{2}+\frac{i\sqrt{3}}{2}$} & 0 & $\frac{9}{4}\omega Y^2$ & $\frac{27}{8}Y^3$ & $0$ & $0$ & $\frac{81}{16}\omega^2 Y^4$ & $0.94867\ldots$ \\  
\multicolumn{1}{l|}{$\tau_{T_{\tau}S_{\tau}}=\frac{1}{2}+\frac{i\sqrt{3}}{2}$} & 0 & $\frac{9}{4}\omega^2Y^2$ & $\frac{27}{8}Y^3$ & $0$ & $0$ & $\frac{81}{16}\omega Y^4$ & $0.94867\ldots$ \\
\multicolumn{1}{l|}{$\tau_{S_{\tau 1}}=i$} & $(6\sqrt{3}-9)Y^2$ & $-(6\sqrt{3}-9)Y^2$ & $0$ & $(189-108\sqrt{3})Y^4$ & $-(189-108\sqrt{3})Y^4$ & $(189-108\sqrt{3})Y^4$ & $1.02253\ldots$ \\
\multicolumn{1}{l|}{$\tau_{S_{\tau 2}}=\frac{3}{2}+\frac{i}{2}$} & $-(6\sqrt{3}+9)Y^2$ & $(6\sqrt{3}+9)Y^2$ & $0$ & $(189+108\sqrt{3})Y^4$ & $-(189+108\sqrt{3})Y^4$ & $(189+108\sqrt{3})Y^4$ & $0.54798\ldots$ \\
\multicolumn{1}{l|}{$\tau_{T_{\tau 1}}=i\infty$} & $Y^2$ & $0$ & $Y^3$ & $Y^4$ & $0$ & $0$ & $1$ \\
\multicolumn{1}{l|}{$\tau_{T_{\tau 2}}=\frac{3}{2}+\frac{i}{2\sqrt{3}}$} & 0 & $Y^2$ & $Y^3$ & $0$ & $0$ & $Y^4$ & $-4.26903\ldots$ \\
\botrule
\end{tabular}}}
\caption{Singlets for the modular forms of weight 2, 4, 6 and 8 of level 3 for four $A_4$ elements.}
\label{table:singlets A4}
\end{table}

For $S_{\tau}:\tau\rightarrow-1/\tau$, and using the stabiliser $\tau_{S_{\tau 1}}=i$, which stabilises the modular transformation represented by the $2\times2$ matrix in Eq.(\ref{generators representation 2times2}), the expression for the modular form at the stabiliser is
\begin{equation}
\rho_{\bft}(S_{\tau})Y_{\bft}^{(2k)}(\tau_{S_{\tau 1}})
=(-\tau_{S_{\tau 1}})^{-2k}Y_{\bft}^{(2k)}(\tau_{S_{\tau 1}})=
(-i)^{-2k}Y_{\bft}^{(2k)}(\tau_{S_{\tau 1}})=
(-1)^{k}Y_{\bft}^{(2k)}(\tau_{S_{\tau 1}}),
\end{equation}
and thus we obtain its directions from the eigenvectors of the representation matrix for S (Eq.(\ref{triplet representations A4})) corresponding to the eigenvalue in the previous equation:
\begin{align}
Y_{\bft}^{(2k)}(\tau_{S_{\tau 1}}) & = y_{\tau_{S_{\tau 1}},1}^{(2k)}
\begin{pmatrix}
1\\ 1 \\ -2
\end{pmatrix} + 
y_{\tau_{S_{\tau 1}},2}^{(2k)}
\begin{pmatrix}
0\\ -1 \\ 1
\end{pmatrix}, ~k=1~(mod~2)\\
Y_{\bft}^{(2k)}(\tau_{S_{\tau 1}}) & = y_{\tau_{S_{\tau 1}}}^{(2k)}
\begin{pmatrix}
1 \\ 1 \\ 1
\end{pmatrix}, ~k=2~(mod~2).
\end{align}

For the lowest weights that appear in TABLE \ref{table:directions A4}, $y_{\tau_{S_{\tau 1}},1}^{(2)}=Y$ and $y_{\tau_{S_{\tau 1}},2}^{(2)}=\sqrt{3} Y$, and from the definitions of the modular forms of higher weight in terms of those of weight 2 we have that $y_{\tau_{S_{\tau 1}}}^{(4)}=(4-2\sqrt{3}) Y^2$ and the factors for the two triplets with weight 6 are obtained similarly.

In models with a single modular symmetry and no other way to break the symmetry (such as flavons), the symmetry is only broken through the modulus and stabilisers can always be considered to be inside the fundamental domain.
When considering multiple moduli, other stabilizers should be considered. Indeed, while it is possible to apply modular transformations to shift one of the modulus to the fundamental domain, it isn't in general guaranteed that the other moduli may be simultaneously shifted to the fundamental domain. Intuitively, this may be understood as the mismatch of the respective residual symmetries.

\section{Models with two modular \texorpdfstring{$A_4$}{A4} symmetries \label{sec:models A4}}

Models that consider the symmetry breaking from multiple modular symmetries groups to a single symmetry group at low energy have already been constructed using $S_4$ symmetry groups in order to obtain TM1 mixing \cite{deMedeirosVarzielas:2019cyj,King:2019vhv}. In this section, we construct a model that considers two $A_4$ modular symmetries in order to obtain TM2 mixing. Although the current experimental evidence excludes TBM mixing, TM2 remains a viable and appealing scheme for lepton mixing.
At high energies, the model is based in two modular symmetries, $A_4^l$ and $A_4^{\nu}$, with modulus fields denoted by $\tau_l$ and $\tau_{\nu}$, respectively. After the modulus fields acquire different VEV's, different mass textures are realised in the charged lepton and neutrino sectors.

\subsection{\texorpdfstring{$A_4^l \times A_4^{\nu} \rightarrow A_4$ breaking}{A4l x A4nu breaking to A4} \label{sec:symmetry breaking A4}}

First, we start by discussing how the symmetry breaking from two independent $A_4^l \times A_4^{\nu}$ to a single $A_4$ is achieved in a general model where the superfields are $L$, which is a doublet of SU(2)$_L$ containing the left-handed leptons and a triplet under $A_4^l$, $\nu^c$, which is a triplet under $A_4^{\nu}$ containing the conjugate of the right-handed neutrino fields added to the Standard Model, and $H_u$, an additional Higgs doublet as required in Supersymmetric models. A bi-triplet $\Phi$, which is a triplet under both $A_4^l$ and $A_4^\nu$, is introduced. Additionally, $Y^\nu$ represents the Yukawa couplings that in the case of modular symmetries should be modular forms. (Our model considers a weight zero modular form, i.e. a modular field independent constant, but it could also be a triplet under $A_4^{\nu}$). 

We consider that neutrinos get their mass through the type I seesaw mechanism and the term from the superpotential that gives rise to a Dirac mass matrix is $\frac{1}{\Lambda} L \Phi Y^\nu \nu^c H_u$. This term is an effective term that can arise from renormalizable interactions of the fields shown with heavy messengers (not shown explicitly - a possibility for the messenger is an electroweak neutral field), see e.g. \cite{deMedeirosVarzielas:2010ppv, deMedeirosVarzielas:2012cet} for UV completions of flavour models.
Considering the multiplication rules for two triplets to get a trivial singlet, the term $\frac{1}{\Lambda} L \Phi Y^\nu \nu^c H_u$ can be explicitly expanded as:
\begin{equation}
\frac{1}{\Lambda}(L_1,L_2,L_3)P_{23}
\begin{pmatrix}
\Phi_{11}&\Phi_{12}&\Phi_{13}\\
\Phi_{21}&\Phi_{22}&\Phi_{23}\\
\Phi_{31}&\Phi_{32}&\Phi_{33}
\end{pmatrix}
P_{23}Y^\nu(\tau_{\nu})\otimes
\begin{pmatrix}
\nu^c_1\\\nu^c_2\\\nu^c_3
\end{pmatrix}H_u,
\end{equation}
where $Y^\nu\otimes\nu^c$ is the product between $Y^\nu$ and $\nu^c$ that gives a triplet of $A_4^{\nu}$, and $P_{23}$ is the matrix that permutes the second and third columns/rows:
\begin{equation}
P_{23}=\begin{pmatrix}
1&0&0\\
0&0&1\\
0&1&0
\end{pmatrix}.
\end{equation}

If $\Phi$ acquires the VEV $\langle\Phi\rangle = v_{\Phi}P_{23}$ (see Appendix \ref{vac_aligns_A4} for more  details), the symmetry $A_4^l\times A_4^{\nu}$ is broken but given that the same transformation $\gamma$ can be performed in $A_4^l$ and $A_4^{\nu}$ simultaneously, there is still a single modular symmetry $A_4$ that is conserved (the diagonal subgroup). Under this symmetry, a modular transformation takes the form
\begin{equation}
\gamma : (\tau_l,\tau_{\nu}) \rightarrow (\gamma \tau_l,\gamma\tau_{\nu}) = \left( \frac{a\tau_l+b}{c\tau_l+d}, \frac{a\tau_{\nu}+b}{c\tau_{\nu}+d} \right), \gamma \in A_4.
\end{equation}
The term $\frac{1}{\Lambda}L\Phi Y^\nu \nu^c H_u$ gets the form $\frac{v_{\Phi}}{\Lambda} (L Y^\nu\nu^c)_\bfs H_u$, which implies a Dirac matrix term for the neutrinos when the Higgs doublet $H_u$ acquires a VEV.

\subsection{Explicit Model}

The model we consider is a model were the Yukawa coupling $Y^\nu$ is simply a constant.
The transformation properties of fields, Yukawa couplings and masses for this model are in TABLE \ref{transformationp A4}. 

\begin{table}[ht]
\centering
\begin{tabular}[t]{c||c|c|c|c|c}
\toprule
Fields & $SU(2)$ & $A_4^l$ & $A_4^{\nu}$ & $2k_l$ & $2k_{\nu}$ \\ 
\colrule 
$L$ & \textbf{2} & $\bft$ & $\bfs$ & $0$ & $-2$ \\ 
$e^c$ & $\bfs$ & $\bfs$ & $\bfs$ & $+6$ & $+2$ \\ 
$\mu^c$ & $\bfs$ & $\bfspp$ & $\bfs$ & $+6$ & $+2$ \\ 
$\tau^c$ & $\bfs$ & $\bfsp$ &  $\bfs$ & $+6$ & $+2$ \\
$\nu^c$ & $\bfs$ & $\bfs$ & $\bft$ & $0$ & $+2$ \\ 
$H_{u,d}$ & \textbf{2} & $\bfs$ & $\bfs$ & $0$ & $0$ \\  
$\Phi$ & $\bfs$ & $\bft$ & $\bft$ & $0$ & $0$ \\
\botrule 
\end{tabular}~~
\begin{tabular}[t]{c||c|c|c|c}
\toprule
Yukawas/Masses & $A_4^l$ & $A_4^{\nu}$ & $2k_l$ & $2k_{\nu}$ \\ 
\colrule
$Y^l$ & $\bft$ & $\bfs$ & $+6$ & $0$ \\ 
$Y^\nu$ & $\bfs$ & $\bfs$ & $0$ & $0$ \\ 
$M_\bfs$ & $\bfs$ & $\bfs$ & $0$ & $+4$ \\ 
$M_\bfsp$ & $\bfs$ & $\bfsp$ & $0$ & $+4$ \\ 
$M_\bft$ & $\bfs$ & $\bft$ & $0$ & $+4$ \\ 
\botrule
\end{tabular}
\caption{Transformation properties of fields, Yukawa couplings and masses for the right-handed neutrinos.}
\label{transformationp A4}
\end{table}

The Yukawa coefficients for the charged leptons are a modular form $Y^l$ which transforms as a triplet of $A_4^l$ with weight $2k_l=+6$, whereas $Y^\nu$ is simply a modulus independent constant, a modular form of weight $0$. For the right-handed neutrino masses we consider three modular forms transforming under $A_4^{\nu}$: $M_{\bfs}$ as a trivial singlet $\bfs$, $M_{\bfsp}$ as a singlet $\bfsp$ and $M_\bft$ as a triplet $\bft$, all with weights $2k_{\nu}=+4$. The weights were chosen in such a way that the modular forms acquire the desired directions as we show below.

The right-handed electron, muon and tau fields are respectively singlets $\bfs$, $\bfspp$ and $\bfsp$ of $A_4^l$ and trivial singlets $\bfs$ of $A_4^{\nu}$, with weights $2k_l=+6$ and $2k_{\nu}=+2$. The lepton doublets $L$ are arranged as a triplet of $A_4^l$ and a singlet of $A_4^{\nu}$, with weights $2k_l=0$ and $2k_{\nu}=-2$. In this model, the three right-handed neutrinos introduced form a triplet of $A_4^{\nu}$ with weight $2k_{\nu}=+2$.
These are the correct choices for the weights such that the modular forms and fields in each term sum up to zero since the weight for the fields is not $+2k$, which are the values that were introduced in this section, but $-2k$ instead (recall the transformation relations for the modular forms and the superfields, Eq.(\ref{transform superfields and couplings for multi fields}), and how the signs of the exponents where the weights enter differ).

Note that, in spite of the charged leptons only having non-trivial singlet transformations under $A_4^l$ and the right-handed neutrinos only under $A_4^\nu$ (which justifies the nomenclature used), the respective weights introduce non-trivial transformations under both modular symmetries for these fields.

With the fields assigned in this manner, the superpotential for this model, which can be separated into one part containing the mass terms for the charged leptons and the other the neutrino mass terms, has the following form: 
\begin{align}
w&=w_e+w_{\nu},\\ 
w_e &= \left( \alpha (LY^l(\tau_l))_\bfs e^c + \beta (LY^l(\tau_l))_\bfsp \mu^c + \gamma (LY^l(\tau_l))_\bfspp \tau^c \right) H_d ,\\
w_{\nu} &= \frac{Y^\nu}{\Lambda} L\Phi \nu^c H_u + 
\frac{1}{2}M_\bfs(\tau_{\nu})(\nu^c \nu^c)_\bfs +
\frac{1}{2}M_\bfsp(\tau_{\nu})(\nu^c \nu^c)_\bfspp +
\frac{1}{2}M_\bft(\tau_{\nu})(\nu^c \nu^c)_\bft.
\end{align}

The bi-triplet $\Phi$ will then acquire a VEV as presented in Section \ref{sec:symmetry breaking A4}, and the two modular symmetries are broken to a single $A_4$, getting for $w_\nu$ (the $w_e$ terms remain exactly the same):
\begin{align}
w_{\nu} &= y_D(L\nu^c)_\bfs H_u + 
\frac{1}{2}M_\bfs(\tau_{\nu})(\nu^c \nu^c)_\bfs +
\frac{1}{2}M_\bfsp(\tau_{\nu})(\nu^c \nu^c)_\bfspp +
\frac{1}{2}M_\bft(\tau_{\nu})(\nu^c \nu^c)_\bft,
\end{align}
where $y_D=Y^\nu v_{\Phi}/\Lambda$.

The flavour structure after $A_4$ symmetry breaking now follows.
We assume that the charged lepton modular field $\tau_l$ acquires the VEV $\langle\tau_l\rangle=\tau_T=\frac{3}{2}+\frac{i}{2\sqrt{3}}$, which is a stabiliser of $T_\tau$. 
At this stabiliser, a residual modular $Z_3^T$ symmetry is preserved in the charged lepton sector. This implies that the modular form $Y^l$, which has weight $+6$, has the direction
\begin{equation}
Y^l(\tau_l) \propto \begin{pmatrix} 1\\0\\0 \end{pmatrix}
\end{equation}
This direction leads to a diagonal charged lepton mass matrix when the Higgs field $H_d$ acquires a VEV $\langle H_d \rangle = (0,v_d)$:
\begin{equation}
m_e=\begin{pmatrix}
\alpha & 0 & 0 \\
0 & \beta & 0 \\
0 & 0 & \gamma
\end{pmatrix}.
\label{mass matrix charged leptons A4}
\end{equation}
The masses for the charged leptons can be reproduced by adjusting the parameters $\alpha$, $\beta$ and $\gamma$. These constants were redefined to include the constant associated with $Y^l(\tau_l)$ and $v_d$.

For the other modular field $\tau_{\nu}$, since we want to obtain the trimaximal mixing TM2, 
which preserves the second column of the tri-bimaximal mixing matrix, the modular form $M_\bft$ should acquire the direction
\begin{equation}
M_\bft(\tau_\nu) \propto 
\begin{pmatrix} 
1\\1\\1 
\end{pmatrix},
\end{equation}
which occurs for the VEV $\langle \tau_{\nu} \rangle = \tau_S = i$, and thus it should have an even $k_\nu$, as happens for $2k_\nu=+4$.
In this case, a residual modular $Z_2^S$ symmetry is preserved in the neutrino sector. 
This implies the following structure for the right-handed neutrino mass matrix:
\begin{equation}
M_R = c_\bfs \begin{pmatrix} 1&0&0 \\ 0&0&1 \\ 0&1&0 \end{pmatrix} +
c_\bfsp \begin{pmatrix} 0&0&1 \\ 0&1&0 \\ 1&0&0 \end{pmatrix} +
\frac{c_\bft}{3} \begin{pmatrix} 2&-1&-1 \\ -1&2&-1 \\-1&-1&2
\end{pmatrix},
\end{equation}
where $c_{\bfs},c_{\bfsp},c_{\bft}$ are complex constants associated with the respective modular form.

The Dirac mass matrix that relates the active and right-handed neutrinos after the Higgs field $H_u$ acquires a VEV $\langle H_u \rangle = (0,v_u)$ is simply
\begin{equation}
M_D=y_D v_u P_{23}.
\end{equation}
We assume here ``left-right" notation even though this matrix is symmetric.

Consequently, the active neutrino mass matrix for the seesaw mechanism gets the form
\begin{equation}
M_{\nu}= - M_D M_R^{-1} M_D^T = - y_D^2 v_u^2 P_{23} M_R^{-1} P_{23}.
\end{equation}

We want now to diagonalize $M_{\nu}$, such that $U^T M_{\nu} U = M_{\nu_d} = \text{diag}(m_1,m_2,m_3)$, where $m_i$ are the neutrino masses and $U$ is an unitary matrix. It is also true that $U^T M_{\nu} U = - U^T M_D M_R^{-1} M_D^T U = M_{\nu_d}$. So $M_D^T U$ also diagonalizes the matrix $M_R^{-1}$ and thus $V = M_D^{-1} U^*$ diagonalizes $M_R$ such that $V^T M_R V  = M_{R_d} = \text{diag}(M_1,M_2,M_3)$ where $M_i= - \frac{y_D^2 v_u^2}{m_i}$. Conversely, $U = M_D^* V^*$ when $V$ diagonalizes $M_R$.

In the present model, when we apply the tri-bimaximal matrix in Eq.(\ref{TBM matrix})
to the heavy neutrino mass matrix,
we obtain: 
\begin{equation}
U_{TBM}^T M_R U_{TBM} = \begin{pmatrix}
a & 0 & c\\
0 & \frac{a - b}{2} + \sqrt{3}c & 0 \\
c & 0 & b
\end{pmatrix},
\label{eq:UMRU1 A4}
\end{equation}
where $a = c_\bft + c_\bfs - \frac{1}{2}c_\bfsp$, $b = c_\bft - c_\bfs + \frac{1}{2} c_\bfsp$ and $c = \frac{\sqrt{3}}{2}c_\bfsp$.
This matrix has only an element on the second row and second column and four elements on the corners that form a $2\times2$ symmetric matrix and so can be put into block diagonal form by permuting the first and second columns and rows. 
Thus, the full matrix can be fully diagonalized adding a matrix $V_r$ that introduces a rotation among the first and third columns. 
This rotation preserves the second column so $M_R$ is diagonalized by a TM2 mixing matrix, since this mixing matrix can be written as the product of the TBM mixing matrix and a rotation on the first and third columns.
For the present model, $M_D$ is only a permutation, so we have that, being $V = U_{TBM} V_r$ the matrix that diagonalizes $M_R$, the matrix that diagonalizes $M_{\nu}$ is $U = P_{23} U_{TBM} V_r$, which can also be written as $U_{TBM} U_r$, where $U_r$ is a rotation between the first and third columns. Using the parametrization
\begin{equation}
U_r = \begin{pmatrix}
\cos\theta e^{i \alpha_1} & 0 & \sin\theta e^{-i \alpha_2} \\
0 & e^{i \alpha_3} & 0 \\
-\sin\theta e^{i \alpha_2} & 0 & \cos\theta e^{- i \alpha_1}
\end{pmatrix},
\end{equation}
which implies that
\begin{equation}
V_r = \begin{pmatrix}
\cos\theta e^{- i \alpha_1} & 0 & \sin\theta e^{i \alpha_2}\\
0 & e^{- i \alpha_3} & 0 \\
\sin\theta e^{- i \alpha_2} & 0 & -\cos\theta e^{i \alpha_1}
\end{pmatrix},
\label{UTM2 for Mnu A4}
\end{equation}
we are then able to diagonalize both $M_{\nu}$ and $M_R$. Here, $\theta$ is the angle that governs the rotation and the three $\alpha_i$ are introduced such that $M_i$ are purely real values.

It is also possible to start from the diagonal matrix $M_{R_d}$ and get $U_{TBM}^T M_R U_{TBM}$. We have that 
\begin{equation}
V_r^* M_{R_d} V_r^{\dagger} =
\begin{pmatrix}
M_1 \cos^2\theta e^{2 i \alpha_1} + M_3 \sin^2\theta e^{-2 i \alpha_2} & 0 & \frac{1}{2} (M_1 e^{i (\alpha_1+\alpha_2)} - M_3 e^{-i(\alpha_1+\alpha_2)})\sin2\theta \\
0 & M_2 e^{2i\alpha_3} & 0\\
\ast & 0 & M_1 \sin^2\theta e^{2 i \alpha_2} + M_3 \cos^2\theta e^{-2 i \alpha_1}
\end{pmatrix}
\label{eq:VMRdV1 A4},
\end{equation}
where an asterisk was used to omit the non-vanishing off diagonal entry of this symmetric matrix. comparing with Eq.(\ref{eq:UMRU1 A4}) we obtain that
$\alpha_3=\frac{1}{2} \arg\left(\frac{a - b}{2} + \sqrt{3}c\right)$ and, more importantly, we get a mass sum rule for $M_i$ that can also be expressed in terms of the active neutrino masses $m_i$:
\begin{equation}
\begin{split}
\frac{1}{m_2} & = -\frac{1}{y_D^2 v_u^2} \left| \frac{a-b}{2} + \sqrt{3} c \right| \\
& = \left| \frac{1}{2m_1} \left( e^{2i\alpha_1}\cos^2\theta-e^{2i\alpha_2}\sin^2\theta+\sqrt{3} e^{i(\alpha_1+\alpha_2)}\sin2\theta \right) \right. \\
& \left. \hspace{15mm} - \frac{1}{2m_3} \left( e^{-2i\alpha_1}\cos^2\theta - e^{-2i\alpha_2}\sin^2\theta + \sqrt{3} e^{-i(\alpha_1+\alpha_2)}\sin2\theta \right) \right|.
\end{split}    
\label{sum_rule_A4}
\end{equation}

The angles and phases from the standard parametrization of the PMNS matrix in \cite{Zyla:2020zbs} can be expressed in terms of the model parameters $\theta$, $\alpha_1$ and $\alpha_2$ using the expressions between the parameters and the PMNS matrix elements (these expressions are equivalent to the ones in \cite{Novichkov:2018yse})
\begin{align}
\label{sintheta13_relation_A4} \sin^2\theta_{13} &= |U_{e3}|^2  = \frac{2\sin^2\theta}{3} \\
\label{sintheta12_relation_A4} \sin^2\theta_{12} &= \frac{|U_{e2}|^2}{1-|U_{e3}|^2} = \frac{1}{3-2\sin^2\theta} \\
\label{sintheta23_relation_A4} \sin^2\theta_{23} &= \frac{|U_{\mu3}|^2}{1-|U_{e3}|^2} = \frac{1}{2} + \frac{\sqrt{3}}{2} \frac{ \sin 2\theta}{2+\cos 2\theta}\cos (\alpha_1-\alpha_2) \\
\delta &= - \arg \left( \frac{ U_{e3} U_{\tau 1} U_{e1}^* U_{\tau 3}^*}{\cos\theta_{12} \sin\theta_{13} \cos^2\theta_{13} \cos\theta_{23}} 
+ \cos\theta_{12} \sin\theta_{13} \cos\theta_{23} \right) \nonumber \\ 
\label{delta_relation_A4} & = \arg \left(\left(e^{i (\alpha_1-\alpha_2)} \sin^2\theta - 3 e^{- i (\alpha_1-\alpha_2)} \cos^2\theta \right) \sin2\theta \right).
\end{align}

Using the $3\sigma$ C.L. range of $\sin^2\theta_{13}$ for NO(IO), $0.02034(0.02053)\rightarrow0.02430(0.02436)$ \cite{Esteban:2020cvm}, we obtain the allowed range for $\sin\theta$:
\begin{equation}
0.1747(0.1755)\lesssim|\sin\theta|\lesssim0.1909(0.1912),
\end{equation} 
which implies also ranges for the other mixing angles  (using that $-1\leq\cos(\alpha_1-\alpha_2)\leq1$): 
\begin{align}
0.3403 (0.3403) \lesssim \sin^2\theta_{12} \lesssim 0.3416 (0.3417) \\
0.3891 (0.3890) \lesssim \sin^2\theta_{23} \lesssim 0.6109 (0.6110).
\end{align}
The $1\sigma$ region is within the interval found for $\sin^2\theta_{23}$, which overlaps with the $3\sigma$ region for this parameter, with our result extending below 0.407(0.411) for NO(IO) and not reaching its upper limit. The range of allowed values for $\sin^2\theta_{12}$ is near the upper allowed limit, which is a characteristic feature of the TM2 mixing, since the lowest value allowed for $\sin^2\theta_{12}$ is $1/3$ as can be seen from Eq.(\ref{sintheta12_relation_A4}). 

The sum rule Eq.(\ref{sum_rule_A4}) and Eqs.(\ref{sintheta13_relation_A4}-\ref{delta_relation_A4}) provide us relations between the six observables and the five parameters of the TM2 mixing, and hence provide what is needed to do a numerical minimization recurring to the $\chi^2$ function:
\begin{equation}
\chi^2 = \sum_i \left(\frac{P_i(\{x\})-BF_i}{\sigma_i}\right)^2,
\end{equation}
where $P_i$ are the values provided by the considered model, $BF$ the best fit value from NuFit \cite{Esteban:2020cvm} and $\sigma_i$ is also provided by NuFit, when averaging the upper and lower $\sigma$ provided.
For the fitting, the three mixing angles, the atmospheric and solar neutrino squared mass differences and the Dirac neutrino CP violation phase were considered. 

The fit parameters obtained for normal ordering (NO) and inverted ordering (IO) of neutrino masses can be found in TABLE \ref{fitresults A4}. The best fit values lie inside the 1$\sigma$ range for all the observables except $\theta_{12}$, for both orderings near the upper limit of the 3$\sigma$ range, and $\delta$ for IO. Nonetheless, all the observables are within their $3\sigma$ intervals. The best-fit occurs for normal ordering of neutrino masses with a $\chi^2=9.41$. 
\begin{table}[htb]
\centering
{\renewcommand{\arraystretch}{1.3}
\begin{tabular}{| c | c | c c c c c c c |}
\cline{1-9}
\multirow{4}{*}{NO} & \multirow{2}{*}{Para.} & & $\chi^2$ & $\theta$ & $\alpha_1$ & $\alpha_2$ & $m_1$ & $m_3$ \\
\cline{3-9}
& & & 9.41 & 10.51\degree & -67.60\degree & -24.26\degree & 0.0141 eV & 0.0521 eV \\
\cline{2-9}
& \multirow{2}{*}{Obs.} & $\theta_{12}$ & $\theta_{23}$ & $\theta_{13}$ & $\delta$ & $\Delta m_{21}^2$ & $\Delta m_{31}^2$ & $m_{\beta\beta}$\\
\cline{3-9}
& & 35.72\degree & 49.4\degree & 8.56\degree & 224\degree & 7.42$\times 10^{-5}$eV$^2$ & 2.514$\times 10^{-3}$eV$^2$ & 0.0131 eV\\
\cline{1-9}
\multirow{4}{*}{IO} &\multirow{2}{*}{Para.} & & $\chi^2$ & $\theta$ & $\alpha_1$ & $\alpha_2$ & $m_1$ & $m_3$ \\
\cline{3-9}
& & & 12.2 & 10.56\degree & -95.56\degree & -38.93\degree & 0.0546 eV & 0.0236 eV \\
\cline{2-9}
& \multirow{2}{*}{Obs.} & $\theta_{12}$ & $\theta_{23}$ & $\theta_{13}$ & $\delta$ & $\Delta m_{21}^2$ & $\Delta m_{32}^2$ & $m_{\beta\beta}$\\
\cline{3-9}
& & 35.73\degree & 48.4\degree & 8.61\degree & 237\degree & 7.42$\times 10^{-5}$eV$^2$ & -2.496$\times 10^{-3}$eV$^2$ & 0.0174 eV \\
\cline{1-9}
\end{tabular}}
\caption{Parameters (Para.) and observables (Obs.) for the best fit point for normal and inverted ordering.}
\label{fitresults A4}
\end{table}

\begin{figure}[hbt]
     \centering
     \begin{subfigure}[t]{0.49\textwidth}
         \centering
         \includegraphics[width=\textwidth]{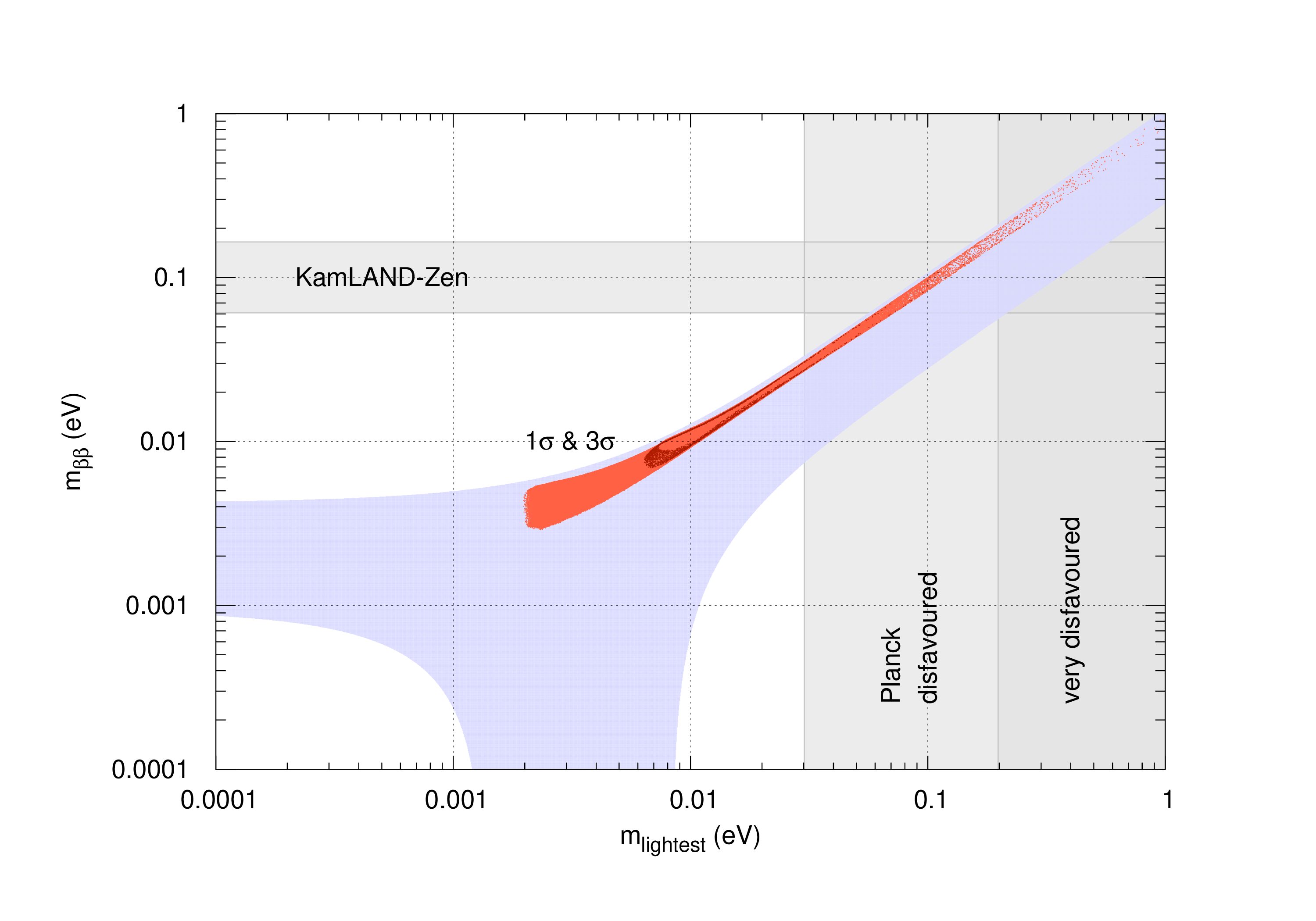}
		 \caption{Normal Ordering}
		 \label{fig:mee NO A4}
     \end{subfigure}%
     ~
     \begin{subfigure}[t]{0.49\textwidth}
         \centering
         \includegraphics[width=\textwidth]{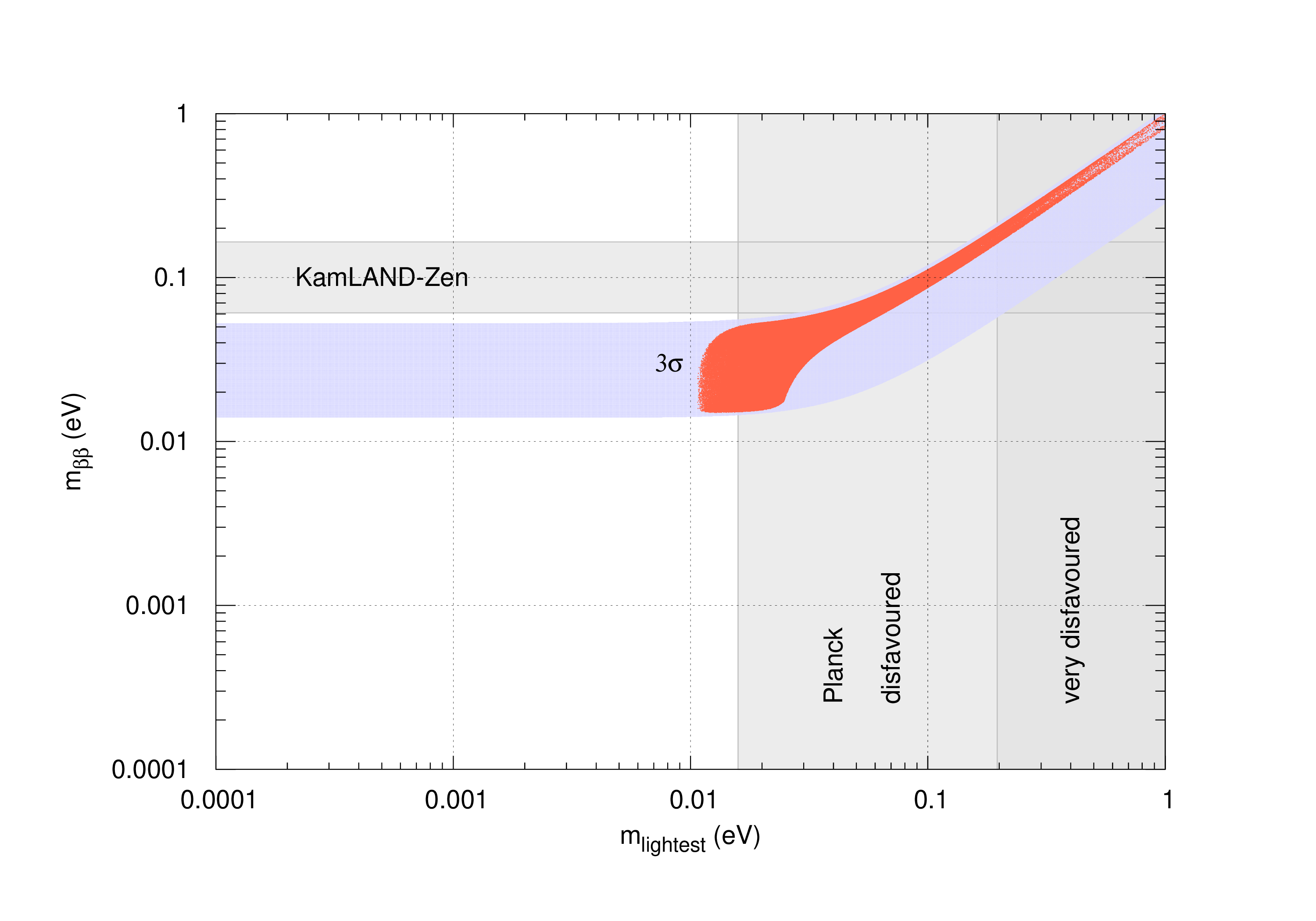}
         \caption{Inverted Ordering}
		 \label{fig:mee IO A4}
     \end{subfigure}
        \caption{Predictions of $m_{\text{lightest}}$ vs $m_{\beta\beta}$ for both ordering of neutrino masses compatible with $1\sigma$ (dark-red, NO only, except $\theta_{12}$) and $3\sigma$ data from \cite{Esteban:2020cvm}. In both figures there were also included the current upper limit from KamLAND-Zen $m_{\beta\beta} < 61-165$ meV \cite{KamLAND-Zen:2016pfg} and cosmological constraints from PLANCK 2018 (disfavoured region 0.12 eV $ <\sum m_i <$ 0.60 eV and very disfavoured region $\sum m_i > 0.60$ eV) \cite{Aghanim:2018eyx}.}
        \label{fig:mee A4}
\end{figure}

It is also possible to obtain the expected $m_{\beta\beta}$ for neutrinoless beta decay using the formula
\begin{align}
m_{\beta\beta} & = \left|(M_{\nu})_{(1,1)}\right| = y_D^2 v_u^2 \left|(M_{R}^{-1})_{(1,1)}\right| \nonumber \\
& = \frac{1}{3} \left| 2 m_1 e^{-2i\alpha_1}\cos^2\theta + m_2 e^{-2i\alpha_3} + 2 m_3 e^{2i\alpha_2}\sin^2\theta \right|,
\label{eq:neutrinoless A4}
\end{align}
where $m_2$ is given by Eq.(\ref{sum_rule_A4}).
Doing a numerical computation, the allowed regions of $m_{\text{lightest}}$ vs $m_{\beta\beta}$ of FIG.\ref{fig:mee A4} (for NO, $m_{\text{lightest}} = m_1$ and for IO, $m_{\text{lightest}} = m_3$) were obtained, using again as constraints the data from \cite{Esteban:2020cvm}. 
In both figures it is also shown the current upper limit provided by KamLAND-Zen, $m_{\beta\beta} < 61-165$ meV \cite{KamLAND-Zen:2016pfg}. Results from PLANCK 2018 constrain the sum of neutrino masses, although different constrains can be obtained depending on the data considered (for more details, see \cite{Aghanim:2018eyx}). In the figures are plotted two shadowed regions, a very disfavoured region $\sum m_i > 0.60$ eV (considering the limit 95\%C.L.,Planck lensing$+$BAO$+\theta_{MC}$) and a disfavoured region $\sum m_i>$ 0.12 eV (considering the limit 95\%C.L.,Planck TT,TE,EE$+$lowE$+$lensing$+$BAO$+\theta_{MC}$). These constraints on $\sum m_i$ can be expressed as constraints on $m_{\text{lightest}}$ using the best fit value for the squared mass differences: $m_{\text{lightest}}>0.198$ eV and $m_{\text{lightest}}>0.030$ eV for NO and $m_{\text{lightest}}>0.196$ eV and $m_{\text{lightest}}>0.016$ eV for IO, for the very disfavoured and the disfavoured regions respectively. We conclude then that only the fit for NO in TABLE \ref{fitresults A4} is outside the disfavoured region.

For NO, there are some points compatible with the $1\sigma$ ranges of the observables other than $\theta_{12}$ (which is, as already said, always near the upper $3\sigma$ limit). These points were plotted with a darker red color. 
For IO, at least one of the other observables is incompatible with its $1\sigma$ region, hence only the $3\sigma$ compatible points are shown for IO. Both mass orderings have points outside the disfavoured region, although the non-disfavoured region for IO is smaller.
The minimum values considering the $3\sigma$ ranges are
\begin{align}
(m_{\text{lightest}})_{\text{min}}^{\text{NO}}\approx  0.002\text{ eV}~~&~~(m_{\beta\beta})_{\text{min}}^{\text{NO}}\approx 0.003 \text{ eV}\nonumber
\\
(m_{\text{lightest}})_{\text{min}}^{\text{IO}}\approx 0.011\text{ eV}~~&~~(m_{\beta\beta})_{\text{min}}^{\text{IO}}\approx 0.015 \text{ eV}.
\end{align}

For this model, NO is hence the preferred mass ordering, although this means smaller values of $m_{\beta\beta}$, which are harder to access experimentally.

\section{Conclusion \label{sec:conclusion}}

We employed the framework of multiple modular symmetries to build a model with minimal field content that is able to reproduce viable Tri-Maximal 2 mixing. We describe how the multiple $A_4$ modular symmetries can break to a single group and show a possible assignment of fields and weights under the modular symmetries leading to the desired mixing scheme. This model is shown to be predictive and to reproduce the observed mixing angles with good fits. Neutrinoless double beta decay is expected, with the inverted ordering possibility almost entirely disfavored by cosmological observations and less compatible with the $1\sigma$ best fit intervals for the experimental observables than the normal ordering of neutrino masses.

\section*{Acknowledgments}
The authors thank João Penedo and Ye-Ling Zhou for helpful discussions.
IdMV acknowledges funding from Funda\c{c}\~{a}o para a Ci\^{e}ncia e a Tecnologia (FCT) through the contract IF/00816/2015 and was supported in part by FCT through projects CFTP-FCT Unit 777 (UID/FIS/00777/2019), PTDC/FIS-PAR/29436/2017, CERN/FIS-PAR/0004/2019 and CERN/FIS-PAR/0008/2019 which are partially funded through POCTI (FEDER), COMPETE, QREN and EU.
JL acknowledges funding from Funda\c{c}\~{a}o para a Ci\^{e}ncia e a Tecnologia (FCT) through project CERN/FIS-PAR/0008/2019.

\appendix

\section{\texorpdfstring{$A_4$}{A4} multiplication rules \label{app:A4}}

The group $A_4$ is the group of even permutations of four objects, see e.g. \cite{Novichkov:2018yse}. It has 12 elements and two generators, S and T:
\begin{equation}
S^2=(ST)^3=T^3=1.
\end{equation}

$A_4$ has four conjugacy classes: $C_1=\{e\}$, $C_2=\{T,ST,TS,STS\}$, $C_3=\{T^2,ST^2,T^2S,TST\}$, $C_4=\{S,T^2ST,TST^2\}$ \cite{Feruglio:2017spp}.

This group has four irreducible representations: an invariant singlet $\bfs$, two non-invariant singlets $\bfsp$ and $\bfspp$, and a triplet $\bft$. The representations for the generators are in TABLE \ref{table:representations A4}.
The three dimensional representation is not determined uniquely but up to an unitary transformation, representing a change of basis. A possible basis is the complex basis, in which $T$ is diagonal, another possible basis is the real basis, in which $S$ is diagonal.

\begin{table}[htb]
\centering
{\renewcommand{\arraystretch}{1.3}
\begin{tabular}{c|c|c|c|c|c} 
\toprule
& ~~$\bfs$~~ & ~~$\bfsp$~~ & ~~$\bfspp$~~ & ~~$\bft$ - complex basis - $\rho$~~ & ~~$\bft$ - real basis - $\tilde{\rho}$~~\\ [0.5ex] 
\colrule
~~$S$~~ & 1 & 1 & 1 & $\frac{1}{3}\begin{pmatrix} -1&2&2\\2&-1&2\\2&2&-1 \end{pmatrix}$~~~ & $\begin{pmatrix} 1&0&0 \\ 0&-1&0 \\ 0&0&-1 \end{pmatrix}$ \\
$T$ & 1 & $\omega$ & $\omega^2$ & $\begin{pmatrix} 1&0&0\\0&\omega&0\\0&0&\omega^2 \end{pmatrix}$ & $\begin{pmatrix} 0&1&0 \\ 0&0&1 \\ 1&0&0 \end{pmatrix}$ \\
\botrule
\end{tabular}}
\caption{Representation for the two generators of $A_4$, where $\omega=e^{i2\pi/3}=-1/2+i\sqrt{3}/2$.}
\label{table:representations A4}
\end{table}

To transform from one basis to the other, we use
\begin{equation}
\tilde{\rho}_\bft(\gamma) = U_\omega \rho_\bft(\gamma) U_\omega^\dagger,
\end{equation}
where the change of basis matrix is
\begin{equation}
U_\omega =\frac{1}{\sqrt{3}}
\begin{pmatrix}
1 & 1 & 1\\
1 & \omega^2 & \omega \\
1 & \omega & \omega^2 \\
\end{pmatrix}
\end{equation}
and obeys $U_\omega^\dagger=U_\omega P_{23}$.

The product of two triplets decomposes as $\bfs+\bfs'+\bfs''+\bft_S+\bft_A$ where $\bft_{S(A)}$ denotes the symmetric (antisymmetric) combination. In the complex basis, this decomposition is \cite{Novichkov:2018yse}
\begin{align}
\begin{pmatrix}
a_1\\
a_2\\
a_3
\end{pmatrix}_\bft
\otimes
\begin{pmatrix}
b_1\\
b_2\\
b_3
\end{pmatrix}_\bft
& = (a_1b_1+a_2b_3+a_3b_2)_\bfs
\oplus(a_3b_3+a_1b_2+a_2b_1)_{\bfs'}
\oplus(a_2b_2+a_3b_1+a_1b_3)_{\bfs''}\nonumber\\
&\oplus\frac{1}{3}
\begin{pmatrix}
2a_1b_1-a_2b_3-a_3b_2\\
2a_3b_3-a_1b_2-a_2b_1\\
2a_2b_2-a_3b_1-a_1b_3
\end{pmatrix}_{\bft_S}
\oplus\frac{1}{2}
\begin{pmatrix}
a_2b_3-a_3b_2\\
a_1b_2-a_2b_1\\
a_3b_1-a_1b_3
\end{pmatrix}_{\bft_A},
\end{align}
and in the real basis it is \cite{Feruglio:2017spp}
\begin{align}
\begin{pmatrix}
a_1\\
a_2\\
a_3
\end{pmatrix}_\bft
\otimes
\begin{pmatrix}
b_1\\
b_2\\
b_3
\end{pmatrix}_\bft
& = (a_1b_1+a_2b_2+a_3b_3)_\bfs
\oplus(a_1b_1+\omega^2a_2b_2+\omega a_3b_3)_{\bfs'}
\oplus(a_1b_1+\omega a_2b_2 + \omega^2 a_3b_3)_{\bfs''}\nonumber\\
&\oplus
\begin{pmatrix}
a_2b_3+a_3b_2\\
a_3b_1+a_1b_3\\
a_1b_2+a_2b_1
\end{pmatrix}_{\bft_S}
\oplus
\begin{pmatrix}
a_2b_3-a_3b_2\\
a_3b_1-a_1b_3\\
a_1b_2-a_2b_1
\end{pmatrix}_{\bft_A}.
\end{align}
Finally, the multiplication rules for the singlets are
\begin{equation}
\bfs\otimes\bfs=\bfs,~~
\bfs'\otimes\bfs'=\bfs'',~~
\bfs''\otimes\bfs''=\bfs',~~
\bfs'\otimes\bfs''=\bfs.
\end{equation}

\section{Modular forms of weight 2 for \texorpdfstring{$A_4$}{A4} \label{modular forms A4}}

The three linearly independent weight 2 modular forms of level 3, $Y_\bft^{(2)}=(Y_1,Y_2,Y_3)$, form a triplet of $A_4$. In \cite{Feruglio:2017spp}, these modular forms were expressed in terms of the Dedekind eta functions
\begin{equation}
\eta(\tau)=q^{1/24}\prod_{n=1}^{\infty}(1-q^n),~~ q=e^{i2\pi\tau}.
\end{equation}

The triplet modular forms $Y_{1,2,3}$ can then be expressed as
\begin{align}
Y_1(\tau)&=\frac{i}{2\pi}\left[ 
\frac{\eta'(\frac{\tau}{3})}{\eta(\frac{\tau}{3})} + 
\frac{\eta'(\frac{\tau+1}{3})}{\eta(\frac{\tau+1}{3})} + 
\frac{\eta'(\frac{\tau+2}{3})}{\eta(\frac{\tau+2}{3})} - 
27\frac{\eta'(3\tau)}{\eta(3\tau)} \right]\\
Y_2(\tau)&=-\frac{i}{\pi}\left[ 
\frac{\eta'(\frac{\tau}{3})}{\eta(\frac{\tau}{3})} + 
\omega^2\frac{\eta'(\frac{\tau+1}{3})}{\eta(\frac{\tau+1}{3})} + 
\omega\frac{\eta'(\frac{\tau+2}{3})}{\eta(\frac{\tau+2}{3})} \right]\\
Y_3(\tau)&=-\frac{i}{\pi}\left[ 
\frac{\eta'(\frac{\tau}{3})}{\eta(\frac{\tau}{3})} + 
\omega\frac{\eta'(\frac{\tau+1}{3})}{\eta(\frac{\tau+1}{3})} + 
\omega^2\frac{\eta'(\frac{\tau+2}{3})}{\eta(\frac{\tau+2}{3})} \right] .
\end{align}
\section{Vacuum alignments for bi-triplet \texorpdfstring{$\Phi$}{Phi} \label{vac_aligns_A4}}

In this Appendix we consider how to align the VEV of the bi-triplet $\Phi$. Following from \cite{deMedeirosVarzielas:2019cyj} where such an alignment was obtained in the context of $S_4$, we add two driving fields, with the properties present in TABLE \ref{tab:driving fields}.

\begin{table}[ht]
\centering
{\renewcommand{\arraystretch}{1.3}
\begin{tabular}{c||c|c|c|c}
\toprule
Fields & $A_4^l$ & $A_4^{\nu}$ & $2k_l$ & $2k_{\nu}$ \\ 
\colrule 
$\chi_{l\nu}$ & $\bft$ & $\bft$ & 0 & 0 \\ 
$\chi_{l}$ & $\bft$ & $\bfs$ & 0 & 0 \\
\botrule 
\end{tabular}}
\caption{Transformation properties of the fields responsible for the vacuum alignment.}
\label{tab:driving fields}
\end{table}

The superpotential responsible for the vacuum alignment that will be minimized with relation to the driving fields is 
\begin{equation}
w=\Phi \Phi \chi_{l\nu} +M\Phi\chi_{l\nu} + \Phi\Phi\chi_l.
\label{eq:superpotential driving fieds}
\end{equation}

Care should be taken given that we are here dealing with $A_4$ groups, rather than $S_4$. The main differences are the presence of the anti-symmetric triplet $\bft_A$ in the contraction of $\bft \times \bft$ (in $S_4$ it is a different inequivalent $\bft'$), and that $S_4$ has a doublet (which decomposes into the two non-trivial singlets of $A_4$). 

As the alignment superpotential above features only contractions into the trivial singlet of $A_4$ and $\Phi \Phi$ contractions (where $\Phi$ appears twice), 
the equations are analogous to those in the $S_4$ case and in general the solutions of these equations are the same as for the $S_4$ case, presented in \cite{deMedeirosVarzielas:2019cyj}. 
Still, the new contraction in $A_4$ that gives a antisymmetric triplet introduces a small difference. When considering the term $\Phi \Phi \chi_l$, we contract $\Phi \Phi$ into a singlet of $A^{\nu}_4$ and a triplet of $A^l_4$ and thus the only non-vanishing contribution is the symmetric triplet of $A^\nu_4$ that is finally combined with $\chi_l$ into a singlet of $A^\nu_4$. Here, no difference appears with relation to $S_4$. 
However, for the term $\Phi \Phi \chi_{\nu l}$, we are now contracting $\Phi \Phi$ into triplets of both symmetries, which means that we will have to consider separately when we contract $\Phi \Phi$ into both symmetric triplets of $A^l_4$ and $A^\nu_4$, and antisymmetric triplets of $A^l_4$ and $A^\nu_4$. The other possibility, i.e. considering simultaneously a symmetric triplet under one symmetry and a antisymmetric triplet under the other, always vanishes.

It is simpler to solve the relations that arise from the minimization of this superpotential working in the real basis. In fact, the multiplication of two triplets in the real basis can be simply expressed by a Levi-Civita tensor.
From Eq.(\ref{eq:superpotential driving fieds}), we have that
\begin{align}
(a \otimes b)_{\bft_{S} i} & = |\epsilon_{ijk}| a_j b_k\\
(a \otimes b)_{\bft_{A} i} & = \epsilon_{ijk} a_j b_k
\end{align}
We get the constraints:
\begin{align}
\sum_{j,k=1,2,3}~\sum_{\beta,\gamma=1,2,3} 
& (g_S |\epsilon_{ijk}||\epsilon_{\alpha\beta\gamma}| + g_A \epsilon_{ijk}\epsilon_{\alpha\beta\gamma})(\tilde{\Phi})_{j\beta}(\tilde{\Phi})_{k\gamma} + M(\tilde{\Phi})_{i\alpha} = 0 ~\text{for}~i=1,2,3, \alpha=1,2,3\\
\sum_{j,k=1,2,3}~\sum_{\alpha=1,2,3} & |\epsilon_{ijk}|(\tilde{\Phi})_{j\alpha}(\tilde{\Phi})_{k\alpha} = 0 ~\text{for}~ i=1,2,3.
\end{align}
where $g_A$ and $g_S$ are constants that account for the combination of both indices of $\Phi \Phi$ symmetrically and anti-symmetrically.
The solutions for general values of $g_S$ and $g_A$, with $g_A \neq g_S$ can be written as $3\times3$ unitary matrices.
\begin{equation}
\begin{split}
\langle\tilde{\Phi}\rangle=v_{\Phi}~ & \left\{
\begin{pmatrix}
1&0&0\\
0&1&0\\
0&0&1
\end{pmatrix},
\begin{pmatrix}
1&0&0\\
0&-1&0\\
0&0&-1
\end{pmatrix},
\begin{pmatrix}
-1&0&0\\
0&-1&0\\
0&0&1
\end{pmatrix},
\begin{pmatrix}
-1&0&0\\
0&1&0\\
0&0&-1
\end{pmatrix}, \right. \\
&\begin{pmatrix}
0&0&1\\
1&0&0\\
0&1&0
\end{pmatrix},
\begin{pmatrix}
0&0&-1\\
-1&0&0\\
0&1&0
\end{pmatrix},
\begin{pmatrix}
0&0&-1\\
1&0&0\\
0&-1&0
\end{pmatrix},
\begin{pmatrix}
0&0&1\\
-1&0&0\\
0&-1&0
\end{pmatrix},\\
&\begin{pmatrix}
0&1&0\\
0&0&1\\
1&0&0
\end{pmatrix},
\begin{pmatrix}
0&1&0\\
0&0&-1\\
-1&0&0
\end{pmatrix},
\begin{pmatrix}
0&-1&0\\
0&0&1\\
-1&0&0
\end{pmatrix},
\begin{pmatrix}
0&-1&0\\
0&0&-1\\
1&0&0
\end{pmatrix},\\
&\begin{pmatrix}
0&0&1\\
0&1&0\\
1&0&0
\end{pmatrix},
\begin{pmatrix}
0&0&-1\\
0&-1&0\\
1&0&0
\end{pmatrix},
\begin{pmatrix}
0&0&-1\\
0&1&0\\
-1&0&0
\end{pmatrix},
\begin{pmatrix}
0&0&1\\
0&-1&0\\
-1&0&0
\end{pmatrix},\\
&\begin{pmatrix}
0&1&0\\
1&0&0\\
0&0&1
\end{pmatrix},
\begin{pmatrix}
0&1&0\\
-1&0&0\\
0&0&-1
\end{pmatrix},
\begin{pmatrix}
0&-1&0\\
1&0&0\\
0&0&-1
\end{pmatrix},
\begin{pmatrix}
0&-1&0\\
-1&0&0\\
0&0&1
\end{pmatrix},\\
& \left. \begin{pmatrix}
1&0&0\\
0&0&1\\
0&1&0
\end{pmatrix},
\begin{pmatrix}
1&0&0\\
0&0&-1\\
0&-1&0
\end{pmatrix},
\begin{pmatrix}
-1&0&0\\
0&0&1\\
0&-1&0
\end{pmatrix},
\begin{pmatrix}
-1&0&0\\
0&0&-1\\
0&1&0
\end{pmatrix}
\right\} .
\end{split}
\label{vev phi solutions}
\end{equation}
where $v_\Phi$ is a constant that depends on $g_A$, $g_S$ and $M$.
These are precisely the representations of the elements of $S_4$ in the real basis, $\tilde{\rho}_\bft(\gamma), \gamma \in S_4$, half of which correspond also to representations of $A_4$ in the real basis. 
Returning to the complex basis used in the main text, we find simply that
\begin{equation}
\langle\Phi\rangle= v_{\Phi} \rho_{\bft}(\gamma) P_{23},~\gamma\in S_4.
\end{equation}

However, in the specific case that $g_A=g_S$, only half of these 24 solutions are valid solutions, more precisely the first twelve solutions in Eq.(\ref{vev phi solutions}), which are the $A_4$ elements in the real basis, and thus,
\begin{equation}
g_A=g_S ~ : ~ \langle\Phi\rangle= v_{\Phi} \rho_{\bft}(\gamma) P_{23},~\gamma\in A_4.
\end{equation}

In the main text we have used as VEV the identity in the real basis, $\delta_{i\alpha}$, first solution in Eq.(\ref{vev phi solutions}), which in the complex basis becomes $\langle\Phi\rangle=v_{\Phi}P_{23}$. 
This specific VEV leads to the recovering of the usual multiplication of two triplets to give a singlet. In the following we will show that it is still possible to construct an invariant term under the single $A_4$ symmetry that remains after the symmetry breaking of the two independent symmetries when choosing one of the other eleven VEV's.

We choose then one of the twelve VEV's $\langle\Phi\rangle=v_{\Phi}\rho_\bft(\gamma_1)P_{23}, ~\gamma_1\in A_4$. We consider that the fields transform under the single $A_4$ as
\begin{align}
E^c & \rightarrow (c_2\tau_l+d_2)^{-2k^l_{E^c}} (c_2\tau_\nu+d_2)^{-2k^\nu_{E^c}} \rho(\gamma_2) E^c\\
L & \rightarrow (c_2\tau_\nu+d_2)^{-2k^\nu_{L}} \rho(\gamma_2) L\\
\nu^c & \rightarrow (c_2\tau_\nu+d_2)^{-2k^\nu_{\nu^c}} \rho(\gamma_1^{-1}\gamma_2\gamma_1) \nu^c\\
Y^l & \rightarrow (c_2\tau_l+d_2)^{2k^l_{Y^l}} \rho(\gamma_2) Y^l\\
Y^\nu & \rightarrow Y^\nu\\
M_\bfs & \rightarrow (c_2\tau_\nu+d_2)^{2k^\nu_{M_\bfs}} M_\bfs\\
M_\bfsp & \rightarrow (c_2\tau_\nu+d_2)^{2k^\nu_{M_\bfsp}} \rho(\gamma_2) M_\bfsp\\
M_\bft & \rightarrow (c_2\tau_\nu+d_2)^{2k^\nu_{M_\bft}} \rho(\gamma_1^{-1}\gamma_2\gamma_1)M_\bft
\end{align}
where $E^c$ stands for $e^c$, $\nu^c$ and $\tau^c$. 
For the singlets, it was taken into account that $\rho(\gamma_1^{-1}\gamma_2\gamma_1)=\rho(\gamma_2)$.
We require here that the triplets $\nu^c$ and $M_\bft$, instead of transforming under $\gamma_2 \in A_4$, transform under the conjugate element of $\gamma_2$, which belongs to $A_4$ if $\gamma_1$ also belongs to $A_4$. 
Obviously for the other twelve solutions that belong to $S_4$ but not to $A_4$ this is not verified.

The transformation rules for $\nu^c$ and $M_\bft$ are equivalent to the following ones:
\begin{align}
\left[ \rho(\gamma_1) \nu^c \right] & \rightarrow (c_2\tau_\nu+d_2)^{-2k^\nu_{\nu^c}} \rho(\gamma_2) \left[ \rho (\gamma_1) \nu^c \right]\\
\left[ \rho(\gamma_1) M_\bft \right] & \rightarrow (c_2\tau_\nu+d_2)^{2k^\nu_{M_\bft}} \rho(\gamma_2) \left[ \rho(\gamma_1) M_\bft \right]
\end{align}
which implies that, with a suitable redefinition of $\nu^c$ and $M_\bft$, we recover the single $A_4$ subgroup under which all the terms after $\Phi$ gains a VEV are invariant. 
In conclusion, we found that, in general, half of the values the VEV of $\Phi$ can have (12 in 24) leads to the same results discussed in the main text and nothing new is left to be said about these other 11 solutions, and interestingly these twelve equivalent solutions are the only possible values for the VEV when $g_A = g_S$.

\end{document}